\documentclass{emulateapj}
\shorttitle{Photoevaporating disks}

\newcommand{\um}{${\rm \mu m}$~}
\newcommand{\mm}{${\rm \mu m}$}

\begin{document}

\title{{\it Spitzer}/MIPS 24 \um Detection of Photoevaporating Protoplanetary Disks}

\author{Zoltan Balog\altaffilmark{1,2}, G. H. Rieke\altaffilmark{1}, Kate Y. L. Su\altaffilmark{1}, James Muzerolle\altaffilmark{1}, Erick T. Young\altaffilmark{1}}

\altaffiltext{1}{Steward Observatory, University of Arizona, 933 N. Cherry Av. Tucson, AZ, 85721; zbalog@as.arizona.edu, grieke@as.arizona.edu, ksu@as.arizona.edu, jamesm@as.arizona.edu,eyoung@as.arizona.edu}
\altaffiltext{2}{on leave from Dept of Optics and Quantum Electronics, University of Szeged, H-6720 Szeged Hungary}

\clearpage

\begin{abstract}

We present 24 \um images of three protoplanetary disks being
photoevaporated around high mass O type stars. These objects have
``cometary'' structure where the dust pulled away from the disk by the
photoevaporating flow is forced away from the O star by photon
pressure on the dust and heating and ionization of the gas. Models of 
the 24 $\mu$m and 8 $\mu$m brightness profiles agree with this hypothesis. 
These models show that the mass-loss rate needed to
sustain such a configuration is in agreement with or somewhat less
than the theoretical predictions for the photoevaporation process.

\end{abstract}


\section{Introduction}

The discovery of ``proplyds'' in Orion \citep[e.g.][]{Odel93,Odel94} let us image
directly protoplanetary disks and study their structure. It also led
to a number of theoretical studies of the photoevaporation of disks by
external radiation fields from neighboring hot stars
\citep{John98,Rich98,Rich00,Mats03,Holl04,Thro05}. Many of these
studies predict behavior that agrees well with the
proplyd properties. The proplyds have been observed intensely, for
example through detection of associated molecular hydrogen
\citep{Chen98} and high resolution measurements in the mid-infrared
\citep{Smith05}.

Deeper infrared observations can provide additional constraints for proplyd
models. At 24 \mm, the resolution of {\it Spitzer} is sufficient to
identify and perhaps resolve such objects at distances up to a few kpc.
At longer wavelengths, the resolution of {\it Spitzer} is too low to
resolve these objects, and they will be lost in the confusion of the
surrounding emission by interstellar material.

The predicted characteristics of an evaporating disk in the $\sim$2 to 50 AU
24 $\mu$m-emitting zone differ substantially among the studies (\citet{John98, Rich00, Holl04, Mats03}. 
However, the calculations do generally agree that the process occurs
on a time scale of $\sim 1 - 3 \times 10^5$ years. As a result, one
might expect to observe a wide range of consequences of disk
evaporation. For stars that have dwelt close to an O star since their
formation, we might find severely truncated disks and possibly reduced
output in the 24 \um band. However, if the stellar velocities within
a young stellar cluster are random, a star might enter the intense
radiation field of an O star and begin the erosion of its disk well
after its formation. A system that has just entered an O star
radiation field is likely to be anomalously bright at 24 \um because
dust grains that would normally be shielded from the ambient
ultraviolet photons in the optically thick disk are being torn from it
by the evaporating gas and exposed to this field. \citet{Rich98}
calculate that the nominal disk emission at 24 \um may be boosted
more than an order of magnitude for some stages in the evaporation
process. This paper describes three systems that are in this dramatic
process of early disk evaporation.

\section{Observations}

Our MIPS observations are taken in scan map mode and typically cover
an area of 0.5-1 square degree. The raw data are converted to
calibrated images with the instrument team Data Analysis Tool (DAT)
\citep{Gordon05}. Point source detection limits are as faint as
$\sim$0.4 mJy, although they can be higher in regions with significant
variations in the infrared background. More details on the methods for
reducing the data (including a description of the IRAC data reduction)
can be found in papers describing the ensemble of infrared sources in
the three clusters that are the subject of this paper: IC 1396
\citep{Sici06}, NGC 2244 (Balog et al. in preparation), and NGC
2264 \citep{Teix06}.

In our full GTO program, we have mapped several regions including a total of
about twenty O stars. Based on a hypothesis that an evaporating disk
might have a cometary appearance at 24 $\mu$m, with a photon-pressure-driven 
tail extending radially outward from the O star, we inspected these twenty
regions. Three clear examples were found, illustrated in Figure 1,
with their positions and integrated flux densities given in Table 1. We
found no additional sources with tails extending in other
directions. A simple and very conservative estimate of the probability
of the three sources being chance alignments can be based on the fact
that their tails are all within 10\arcdeg\,of radial. This calculation
would imply a probability of only about 10$^{-4}$ that the alignments
are random. This estimate ignores the unique morphology of these three
sources, so the likelihood is even
lower that they are randomly occurring phenomena.

\begin{figure}
\plotone{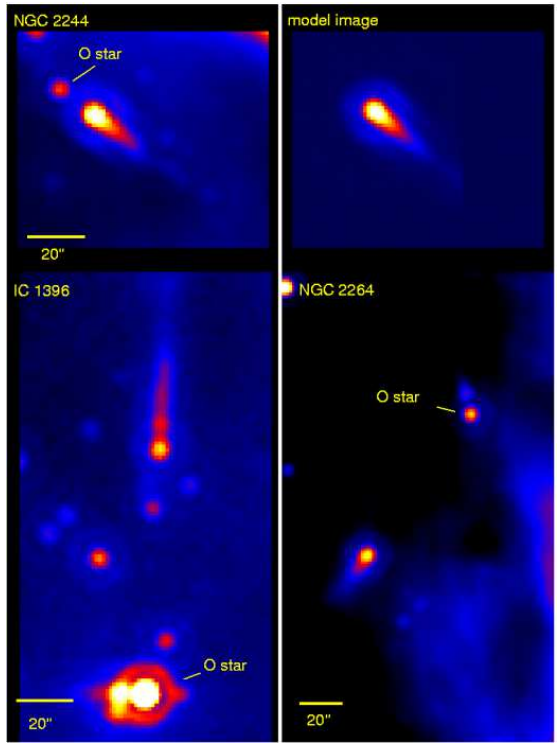} 
\caption{The three evaporating protoplanetary disks together with the
  modeled 24 $\mu$m image for the source in NGC 2244. The orientation
  of each image is north up and east toward the left. The upper right panel is the model 24 \um image for the tail in NGC 2244.}
\label{fig:evap1}
\end{figure}

The position of the head of each ``comet'' coincides with a 2MASS and IRAC 
point source. The sizes of the objects are similar to the cometary clouds of 
photoevaporating globulae found in Carina by \citet{Smith04} but the near IR 
point sources at the positions of the ``comet'' heads indicate that they contain
stars.  No extended structures are observed at wavelengths shorter 
than 24 $\mu$m. The only exception is the source in NGC2244 where the 8 $\mu$m 
image also shows a similar cometary feature. The brightness (Table 1) and the 
infrared 
color of the near infrared objects in the heads indicate that they are stars 
with masses around 0.5 M$_{\odot}$ - 1.0 M$_{\odot}$. 
These mass estimates probably have large errors because 
a significant spread of ages is measured in each cluster.

Table 1. shows the basic parameters of the systems: the cluster to which they belong, the position of the head, the 2MASS K magnitude of the source in the head, the ionizing source and its age, the measured 8 and 24 $\mu$m flux densities including the tail, the length of the tail, the calculated total dust mass and the mass loss rate.

\begin{deluxetable*}{clcccccccc}
\setlength{\tabcolsep}{0.03in}
\tablecolumns{10}
\tabletypesize{\scriptsize}
\tablewidth{0pc}
\tablecaption{The physical and model parameters of the evaporating
  disks and tails}
\tablehead{
\colhead{Cluster}&\colhead{Position} & \colhead{K}& \colhead{inoizing}&\colhead{age$^{\star}$}&\colhead{$F_{8}$}&\colhead{$F_{24}$}
& \colhead{$\Delta l$} & \colhead{$M_{d}^{\dagger}$} &\colhead{$\dot{M}^{\ddagger}$}\\
\colhead{} & \colhead{[J2000.0]} & \colhead{[mag]}&\colhead{source}&\colhead{[Myr]}&\colhead{[mJy]} & \colhead{[mJy]} &\colhead{[pc]} &\colhead{M$_{\sun}$}&\colhead{M$_{\sun}/yr$}
}
\startdata
NGC 2244 & 06:31:54.68 04:56:25.0&13.53&HD46150&4& 1.5$\pm$0.2 & 77$\pm$10 & 0.22 & 3$\times10^{-8}- $ 9$\times10^{-6}$ &  1.4$\times10^{-9}-$ 4.4 $\times 10^{-7}$\\
IC 1396  & 21:38:57.09 57:30.46.5& 13.18&HD206267&3& 2.3$\pm$0.1 & 23$\pm$5  & 0.21  &5$\times10^{-9}- $ 2$\times10^{-6}$ &  2.4$\times10^{-10}-$ 7.7 $\times 10^{-8}$\\
NGC 2264 & 06:41:01.92 09:52:39.0& 13.65 &S Mon&4&2.5$\pm$0.1 & 78$\pm$10 & 0.12 &4$\times10^{-8}- $ 1$\times10^{-5}$ &  3.4$\times10^{-9}-$  1.1 $\times 10^{-6}$\\
\enddata
\tablenotetext{$\star$}{HD46150 \citep{Ogur81},HD206267 \citep{Hesk85}, S Mon \citep{Walk56}}
\tablenotetext{$\dagger$}{Dust mass ranges from the  observed dust  mass using small ($\sim$0.01 \mm) grains based on the 24 \um profile (lower bound) and total mass integrating up to 1 mm size (upper bound).}
\tablenotetext{$\ddagger$}{Mass loss rate is computed assuming a velocity of 10km/s and a gas-to-dust mass ratio of 100.}
\end{deluxetable*}

\section{Analysis and Modeling}

Our analysis is based on the hypothesis that the ``cometary'' tails
represent dust entrained in gas evaporating from the working surface
of the protoplanetary disk and being ejected away from the nearby
O-type star. \citet{Rich98} show theoretical
calculations of the early stages of the process that are compatible
with this possibility. Once the dust has been pulled from the disk, the
smallest particles (radii $\ll$0.01$\mu$m) will be stochastically
heated and some may be destroyed quickly. 
Small particles (radii $<$ 1000 $\mu$m) will be blown
away from the O star by photon pressure, or dragged away by the gas and will 
re-emit the absorbed UV energy in the thermal infrared. 

The mid-infrared brightnesses of our three objects are similar to
those of the Orion proplyds, corrected for distance
\citep{Smith05}. Only the system in NGC 2244 shows slightly extended structure
at 8 \mm; the other two are point-like in the 8 \um images. The fact
that all the tails are much brighter at 24 \um suggests that small
grains (size less than 0.1 \mm) are the dominant emitters, assuming
that the grains reach their equilibrium temperatures. To understand the 
structure of these tails, cuts
with widths of 6\farcs23 were made at 8 and 24$\mu$m along the symmetric axis of the
tail centered at the O star. The background values were estimated
using the area above and below the tail and subtracted off. 
The surface brightness profiles are displayed in Figures
\ref{prof_ngc2244}, \ref{prof_ic1396} and \ref{prof_ngc2264}.

\subsection{Basic Model and Parameters} 

To model the emission of the cometary tails we use the same model
that \citet{Su05} used for the extended Vega debris disk. This model
considers the behavior of dust grains being
blown away from a luminous central star by radiation pressure to form
an extended outflow disk. It should apply generally for outward flows
with a constant number density radially
($n(r)\sim$ constant). To adapt the model to the current situation,
we assume that the cometary outflow emerges from the disk (comet
``head'') position. The remaining parameters are the dust properties 
(composition and grain sizes), the outer radius of the tail, and the angle 
that the line joining the O star and the head along the tail makes with 
respect to the plane of the sky
(0\arcdeg~means the tail is pointed away from the O star exactly on
the plane of sky). The emission from the tail is computed based
on the equilibrium dust temperatures assuming the tail is optically
thin (no stochastic heating is included). A series of model images
was computed using different parameters, convolved with the {\it
Spitzer} PSFs to match the observed resolutions, and then we extracted
radial cuts for comparisons.

We first explored how the observed profiles of the tails vary with
different grain sizes and angles to the plane of the sky. To simplify
the modeling, we assumed the grains are astronomical silicates and
have a uniform size (radius of $a$). The modeled 24 \um profiles
basically fade rapidly as the grain sizes increase. This is because
large grains have colder temperatures than small ones if they are at
the same distance from the radiating source; as a result, the expected 
emission at 24 \um drops
significantly, making a shorter observable tail. A similar
characteristic is seen while varying the angles with respect to the
plane of the sky, because even though the projected distance (on the
plane of the sky) remains the same, the actual distance between the
tail and the O star increases as the angle increases, resulting in
colder dust temperatures and a shorter observable tail. In general,
this simple model is degenerate between the sizes of grains and the
actual angles of the tail with respect to the plane of the sky.
However for the three systems discussed here, models based on grains larger than
$\sim$1 \um come short of the observed tail lengths, and hence cannot fit the 
data even assuming the tails
are on the plane of the sky. Without other data to constrain the model parameters,
hereafter we assume the observed tails are exactly on the plane of the
sky, for simplicity. This assumption is reasonable given the selection
bias that they would be more difficult to recognize at other angles.

\begin{figure}
\plotone{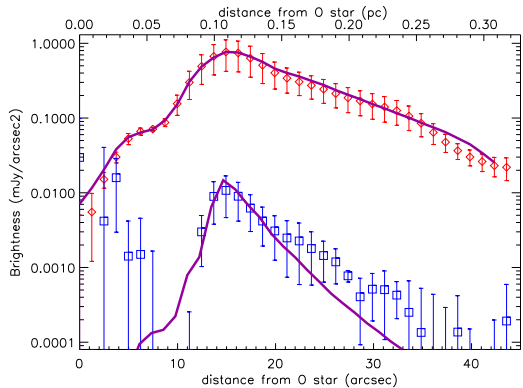}
\caption{The brightness distribution of the tail in NGC 2244 with 
  diamonds for the data at 24 \um and squares at 8 \mm,
  centered at the position of the 05 V star. The contribution of
  the O star has been taken out by PSF subtraction. 
  The 8 \um profile within 8\arcsec ~from the O star is contaminated by the
  residual of the O star. 
  The best-fit model is shown with the solid lines.}
\label{prof_ngc2244}
\end{figure}

\begin{figure}
\plotone{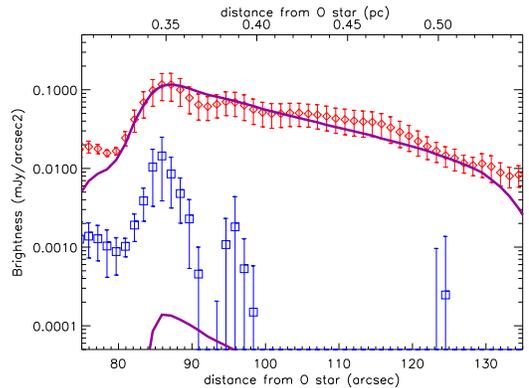}
\caption{The brightness distribution of the tail in IC 1396. Symbols used are 
the same as Figure \ref{prof_ngc2244}.}
\label{prof_ic1396}
\end{figure}

{\it The Tail in NGC 2244:} The tail in NGC 2244
is powered by the O5~V star HD46150, which we represent with a blackbody of 42,000
K and 12 R$_{\sun}$ located at a distance of 1.5 kpc. As
shown in Figure \ref{fig:evap1}, among the three systems this tail is
the closest to the O star with a projected distance of $\sim$0.1 pc
(therefore the warmest). A tail composed of $a$=0.03 \um grains
produces a simultaneously good fit to the profiles at 8 and 24
\mm. Models with other grain sizes can also produce good
fits to the 24 \um profile, but the expected emission at 8 \um is
either much lower than the observed flux (for $0.03<a<0.3$),
or much higher for ($a <0.03$ \mm). The best fit profiles are shown in
Figure \ref{prof_ngc2244} as solid lines. The tail extends from 0.1 to 0.32 pc ($\Delta l \sim 0.22$ pc) from the
O star, with a total observed dust mass of $\sim$3$\times10^{-8}$
M$_{\sun}$. The total model emission of the whole system is $\sim$74
mJy at 24 \um and $\sim$0.5 mJy at 8 \mm.

{\it The Tail in IC 1396:} The distance of IC 1396 is $\sim$835 pc,
much closer than NGC 2244. However, the projected distance between the
disk and the O6~V star HD206267 is the largest (0.35 pc) with the
longest observed angular tail length (0.35-0.56 pc, $\Delta l \sim
0.21$ pc) among the three systems. The tail also appears be clumpy (see
Figure \ref{fig:evap1}), although this could also be due to a chance
alignment with an unrelated source. As in the NGC 2244 object, a tail
composed of $a \lesssim$0.01 \um grains can provide satisfactory fits
to the observed 24 \um profile, but it cannot reproduce the flux at 8
\mm. One possible explanation is there exists a population of very
small grains ($a \sim$10 \AA) in the disk that are stochastically
heated; therefore, the resultant emission spectrum would shift
to shorter wavelengths producing more 8 \um flux than the emission
spectrum using the thermal equilibrium temperature (e.g., Figure 8.12
in \citealt{Krug03}). The dust mass contribution from this very small
grain population should be small. Therefore, the dust mass derived
from our simple model, $\sim$5$\times10^{-8}$ M$_{\sun}$, is a
reasonable estimate. The total model emission at 24 \um is $\sim$19 mJy and 
$<$0.01 mJy at 8 \mm.

\begin{figure}
\plotone{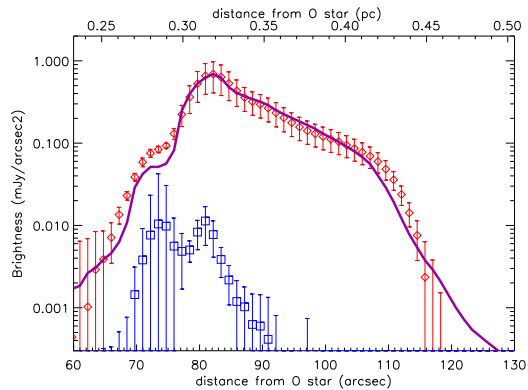}
\caption{The brightness distribution of the tail in NGC 2264. Symbols
  used are the same as Figure \ref{prof_ngc2244}.}
\label{prof_ngc2264}
\end{figure}

{\it The Tail in NGC 2264:} 
The cluster NGC 2264 is located $\sim$800 pc away, similar to IC
1396. The star powering the tail, S Mon, is slightly cooler than the other two
cases, O7~Ve; therefore a blackbody of 38,000 K and 9
R$_{\sun}$ was used. The disk is also farther away from the O star,
0.3 pc, with the shortest tail length (0.32-0.44 pc, $\Delta l \sim
0.12$ pc) among these three systems. In addition, the contrast between
the head (profile within 6\arcsec~from the head position) and
the tail is much larger than the other two cases (see Figure
\ref{prof_ngc2264}). Therefore, similar models used to fit the tails
in NGC 2244 and IC 1396 cannot obtain a good fit to the observed 24
\um profile. An additional unresolved point source at the position of
the disk is added to fit the profile. Our models are also too faint at 8 \mm. The behavior at both wavelengths suggests the presence of a population of very small grains that are rapidly destroyed. The models with grains with $a <$0.01 \um can produce good fits
to the rest of the 24 \um profile. The total observed dust mass is
$\sim 4\times10^{-8}$ M$_{\sun}$, and the total emission at 24 \um is
$\sim$66 mJy and $\ll$0.1 mJy at 8 \mm.

\subsection{Implications} 

Although small grains ($a\sim$0.01 \mm) are responsible for most of
the emission at 24 \mm, larger ``invisible'' grains may contribute
significantly to the total dust mass in the tail. To estimate the
mass, we assume that the grains have a size distribution similar to the
interstellar medium, $N(a) \propto a^{-3.5}$, as is also 
expected for grains undergoing collisional cascades in
protoplanetary disks \citep{Dohn69,Tana96}.
The total dust mass in the tail is then $M_{tot} \propto \int {4 \over
3} \pi a^3 N(a) da \propto \int a^{-0.5} da \propto a^{0.5}$. Integrating
between 0.01 $\mu$m and 1 mm (roughly the largest grain size that can
be ejected by the radiation pressure of an O star) gives an upper limit to
the total dust mass in the tail; values are listed in Table 1. 

We can also estimate the mass loss rate in the photoevaporating disks
given the velocity of the flow in the cometary tails.  Following the
equation given in \citet{Su05}, the terminal velocity ranges from
$\sim$10 to 100 km s$^{-1}$ for a typical O-type star (luminosity of
$10^6$L$_{\sun}$ and mass of 50M$_{\sun}$) and grains with sizes of
1-0.01 \mm, assuming negligible gas drag effects. The presence of gas in the 
tail will result in lower velocities. However, heating and ionization  
accelerate gas to speeds of 3-20 km s$^{-1}$ \citep{John98,Bally98} and 
these flows entrain dust. 

Although the flow velocity is hard to determine accurately, our estimate for 
the mass loss rate is robust because it depends on the product of the velocity and the 
gas to dust ratio which are roughly inversely correlated to each other. For a 
normal gas to dust ratio (100:1) the bulk velocity of the dust and gas mixture 
is on the order of 10 km s$^{-1}$. If the gas to dust ratio deviates from the 
standard value (the effect of gas decreases) the bulk velocity will approach 
the terminal velocity. Although the outflow velocity of material will increase 
the total calculated mass loss will remain approximately the same since the 
lower gas to dust ratio will result in lower total mass, hence a lower mass loss rate.

The X-ray observations of these O stars indicate significant stellar
winds (L$_x$=10$^{32.4}$erg/s for S Mon, \citealt{Simon05};
L$_x$=10$^{32.5}$erg/s for HD46150, \citealt{Walborn02}; and a bright
ROSAT source for HD206267, \citealt{Berghofer96}). The strong stellar
winds would speed up the outflow from the disk if the disk was very close
to the O star (less than 0.01 pc). However, such winds should be only a 
secondary effect at the distances of these objects from the stars.

We can place lower and upper limits on the mass loss rate, corresponding to the observed grain sizes and to the full size range up to 1000 \um. We take a typical gas-to-dust
mass ratio of 100 and a velocity of 10 km/s. The range of mass loss rate is
$1.4\times10^{-10}$ to $4.4\times10^{-8}$ M$_{\sun} yr^{-1}$,
$2.4\times10^{-11}$ to $7.7\times10^{-9}$ M$_{\sun} yr^{-1}$, and
$3.4\times10^{-10}$ to $1.1\times10^{-7}$ M$_{\sun} yr^{-1}$ for the tails
in NGC 2244, IC 1396 and NGC 2264, respectively. To first order, the
upper limits of the derived mass loss rates are somewhat less than the 
predictions by the photoevaporation models,
$\sim1-6\times 10^{-7}$ M$_{\odot} yr^{-1}$ at 0.3 - 0.1 pc
from the UV energy source \citep{Rich98,York04}, and the lower limits
are much less than these expectations.

\section{Conclusions}

We report the detection of three photoevaporating disks with cometary
tails around high-mass stars in {\it Spitzer}/24 $\mu$m images. Their
observed brightnesses and morphologies agree well with a model
where the dust is entrained in the evaporating gas and pulled from the
disk. Once exposed to the O star radiation field, a structure like a
comet tail grows, with dust and gas streaming away from the O star
under the influence of its radiation pressure on the dust and heating
and ionization of the gas. We have roughly
estimated the evaporation mass loss rate from our observations and
find an upper limit in reasonable agreement with or somewhat lower than 
theoretical
expectations. Further observations of these systems should be able to
improve our understanding of the evaporation process substantially.

\acknowledgments

We thank the referee, John Bally for suggestions
 that greatly improved the manuscript.
This work is based on observations made with the Spitzer Space
Telescope, which is operated by the Jet Propulsion Laboratory,
California Institute of Technology, under NASA contract 1407. Support
for this work was provided by NASA through contract 1255094, issued by
JPL/Caltech. ZB received support from Hungarian OTKA Grants TS049872,
T042509 and T049082.

\end{document}